\documentclass[aps,prr,superscriptaddress,twocolumn%,longbibliography
]{revtex4-1}
\usepackage{graphicx}
\usepackage{amsmath}
\usepackage{mathtools}
\usepackage{amsfonts}
\usepackage{bm}
\usepackage[normalem]{ulem}
\usepackage[usenames, dvipsnames]{xcolor}
\usepackage[export]{adjustbox}

\usepackage{hyperref}
\hypersetup{colorlinks=true, linkcolor=blue, citecolor=blue, urlcolor=blue}

\newcommand\blue[1]{{\color{black}#1}}

\begin{document}

\title{Thin-film aspects of superconducting nickelates}

\author{F. Bernardini}
\affiliation{Dipartimento di Fisica, Universit\`a di Cagliari, IT-09042 Monserrato, Italy}
\author{L. Iglesias}
\affiliation{Unité Mixte de Physique, CNRS, Thales, Université Paris-Saclay, 91767 Palaiseau, France}
\author{M. Bibes}
\affiliation{Unité Mixte de Physique, CNRS, Thales, Université Paris-Saclay, 91767 Palaiseau, France}
\author{A. Cano}
\affiliation{
Univ. Grenoble Alpes, CNRS, Grenoble INP, Institut Néel, 25 Rue des Martyrs, 38042, Grenoble, France
}

\date{\today}

\begin{abstract}
The discovery of superconductivity in infinite-layer nickelates has attracted much attention due to their association to the high-$T_c$ cuprates. 
Cuprate superconductivity was first demonstrated in bulk samples and subsequently in thin films. In the nickelates, however, the situation has been reversed: although surging as a bulk phenomenon, nickelate superconductivity has only been reported in thin films so far. At the same time, the specifics of infinite-layer nickelates yield distinct interface and surface effects that determine their bulk vs thin-film behavior. In this paper, we provide an overview on these important aspects. 

\end{abstract}

\maketitle

\section{Introduction}

The infinite-layer nickelates $R$NiO$_2$ ($R=$ rare-earth element) have long been discussed as potential cuprate-like high-$T_c$ superconductors \cite{hayward99,anisimov99,pickett-prb04,norman20-p}. 
This idea can now be scrutinized experimentally after the discovery of superconductivity in hole-doped NdNiO$_2$ \cite{hwang19a} and its subsequent verification in hole-doped  PrNiO$_2$ and LaNiO$_2$ as well \cite{ariando20,hwang20Pr-a,hwang20Pr-b,hwang21-La,ariando21-La,ariando21-interfacial,zhu21-substrates,wen20a}. To date, however, this breakthrough remains limited to thin films (see e.g. \cite{cano-review,tao-review,arita21review}). 
This circumstance is apparently related to the thermodynamic fragility 
of these special phases, in which the interesting electronic properties require an unfavorably low valence of the nickel atom. 
Thus, the simultaneous control of both sample quality and doping necessary to promote superconductivity 
in these nickelates 
turns out to be a real experimental challenge. 
%\textcolor{orange}{Lucia' alternative: This circumstance is apparently related to the thermodynamic instability of these spacial phases, becoming a real experimental challenge the simultaneous control of both sample quality and doping necessary to promote superconductivity in these low-valence nickelates.}
In this respect, the thin-film approach has proven its advantages. 
At the same time, superconductivity has been reported for thicknesses as large as $17$~nm (i.e. $\sim$~50 unit cells), with relatively large critical currents ($\gtrsim$ 200 kA/cm$^2$), and also for thin films on different substrates. Consequently, there is consensus in that nickelate superconductivity is a genuine bulk phenomenon.

The verification and further investigation of nickelate superconductivity in actual bulk samples (ideally single-crystals) has therefore emerged as an important goal in the field. 
In parallel, the efforts made along the thin-film route have very recently enabled the observation of nickelate superconductivity in a quintuple layer system that is hardly realizable in its bulk form \cite{mundy21}.
Thus, the thin-film approach continues to be the favoured option to gain further insight about nickelate superconductivity. At the same time, the specific effects that may be at play in thin-film vs bulk samples need to be better understood. In the following we provide an overview of the current research along this line. 

\section{Epitaxial-growth aspects}

The synthesis of rare-earth infinite-layer nickelate thin films requires the epitaxial growth the perovskite precursors $R$NiO$_3$ in the first place. These perovskite nickelates have been grown by various techniques. The most popular is pulsed laser deposition (PLD) \cite{preziosi_reproducibility_2017,guo_tunable_2020,frano_layer_2014,liu_heterointerface_2013}, but sputtering \cite{scherwitzl_electric-field_2010,mikheev_tuning_2015} and molecular beam epitaxy \cite{king_atomic-scale_2014} have also been used. At this stage, the main difficulty lies in the stabilization of the unfavorable high 3+ valence of Ni (instead of its most preferred 2+)\blue{, reproducibility and off-stoichiometry issues have been reported in the literature \cite{bibes17}}. 
This difficulty increases for Ca- or Sr-doped films, as required to obtain the subsequent infinite-layer phases in the appropriate regime of Ni 3$d$-electron filling for superconductivity (i.e. $\sim d^{8.8}$). 
In PLD this is typically achieved by starting from a mixed-phase polycrystalline target and using a highly oxygen rich atmosphere (on the order of 0.15-0.3 mbar), temperature around 600-650\,$^{\circ}$C  and a laser fluence ranging from 0.9 to 2~J\,cm$^{-2}$. It was also shown that parameters such as the laser spot size or the target history could also play an important role in obtaining single phase perovskite films.
However, the synthesis of hole doped nickelates is particularly challenging, not only due to the instability of high-valence Ni itself but also because of the competition between the perovskite phase and Ruddelsden-Popper phases \cite{hwang20-apl,zhu21-cpl}. 
These two factors are the main obstacles to obtain single-phase perovskite films.
%but the maximization of 
The fraction of perovskite phase present in the film can be optimized by monitoring the intensity of the (001) diffraction peak (which is absent in the Ruddelsden-Popper compound) and the position of the (002) peak, which exists in both the perovskite and Ruddelsden-Popper phases but appears at a larger angle for the perovskite (above 48$^{\circ}$). Lee et al \cite{LeeAPL2020} have shown that to optimize both criteria it is preferable to work with relatively high laser fluence and very small laser spot size (on the order of 2 mm$^2$ or less).

Even if single-phase perovskite thin films are obtained, it is important to keep in mind that the quality of the samples may be limited by the presence of additional atoms or defects. At the interfaces, in particular, epitaxial strain may favor the presence of oxygen or even $A$-site vacancies (see e.g. \cite{SpaldinPRB2013, Petrie2016}). What is more, achieving a single termination is difficult in many substrates and different interfacial configurations are generally possible in these systems (see e.g. \cite{ahn14-pra,fong21-apl}).  
Consequently, the Ni atom may be embedded in multiple interfacial environments already at this stage.

After the precursor perovskite phase has been obtained, the samples are reduced by thermal annealing in a vacuum-sealed tube containing a small amount of a highly reducing agent (CaH$_2$). The goal is to selectively remove one-third of oxygens (i.e., one full plane of apical oxygens) and thereby stabilize the oxygen poorer infinite-layer phase. The literature diverges regarding the exact conditions for this process, but it typically has to be performed between 200 and 360 $^{\circ}$C for several hours \cite{LeeAPL2020, zhu21-cpl, ariando20, hwang20Pr-a, ariando21-La}. In the ideal case, X-ray diffraction then reveals the complete transformation of the perovskite phase into the infinite-layer phase, with a characteristic shift of the (002) peak position to higher angles ($\sim$54$^{\circ}$) signalling the contraction of the out-of-plane lattice constant, an intense (001) peak (its absence or reduced intensity is again the signature of spurious phases) and Laue fringes, attesting of the good structural coherence of the infinite-layer phase in the growth direction \cite{LeeAPL2020, zhu21-cpl, hwang21-La}.

\section{Epitaxial strain}

The most obvious additional ingredient that appears in thin films is the epitaxial ---or biaxial--- strain associated to the lattice mismatch between the nickelate and the substrate. 
%\textcolor{orange}{Lucia alternative: In general terms, imposing an in-plane compressive (tensile) strain results in out-of-plane expansion (contraction) and the epitaxial strain is defined as $(a_\text{substrate} - a)/a$. For instance, the lattice parameters of bulk LaNiO$_2$ have been reported to be $a = 3.96$~{\AA} and $c = 3.37$~{\AA} \cite{hayward99}. Thus, when epitaxial thin films are grown on the (001) SrTiO$_3$ substrate, the in-plane $a$ lattice parameter is forced to match to the 3.905~{\AA} of SrTiO$_3$, which leads to an increase of the out-of-plane $c$ lattice parameter (since the system is only free to relax the stress in that direction).  Consequently, the epitaxial strain in the case of LaNiO$_2$ is $-$1.4~\%.}
For instance, the lattice parameter of the common substrate SrTiO$_3$ is 3.905~{\AA} while the lattice parameters of bulk LaNiO$_2$ have been reported to be $a = 3.96$~{\AA} and $c = 3.37$~{\AA} \cite{hayward99}.  
Thus, for $c$-axis oriented thin films of LaNiO$_2$ grown on SrTiO$_3$(001) the epitaxial strain, which is defined as $(a_\text{substrate} - a)/a$, will be $-$1.4~\%.  
%when epitaxial thin films are grown on the (001) SrTiO$_3$ substrate with the $c$ axis perpendicular to its (001) direction, then the $a$ parameter is forced to match 3.905~{\AA} at the interface. 
%Consequently, the epitaxial strain defined as $(a_\text{substrate} - a)/a$ will be $-$1.4~\%.  
In general, this reduction of the $a$ parameter is accompanied with an increase in $c$ since the system is free to relax the stress in that direction. That is, imposing an in-plane compression at the interface results in out-of-plane tensile strain. 
Conversely, in-plane tensile strain ---due to a different substrate for example--- can be expected to produce an out-of-plane compression.

By means of these distortions of the cell parameters, epitaxial strain can further modify the key features of the corresponding electronic structure (see e.g. \cite{botana20magnetism}). The ``cuprateness'' of this structure, in particular, is generally defined from the nature of the states at the Fermi level and the so-called charge-transfer energy (see e.g. \cite{pickett-prb04,botana20prx,cano20b,kotliar20prb,lechermann20prb}).
In this respect, in-plane compressive strain increases both the bandwidth of the main Ni-3$d_{x^2-y^2}$ band crossing the Fermi level and its self-doping with the $R$-$5d$ states. At the same time, the O-2$p$ states are pushed further below the Fermi level, thereby increasing the corresponding charge-transfer energy. These changes that can be obtained locally due to epitaxial strain have been argued to mimic the trends across the $R$NiO$_2$ series, including the corresponding tendency towards magnetic order \cite{botana20magnetism}. Furthermore, epitaxial strain can also induce changes in the crystal structure itself and promote a $P4/mmm \to I4/mcm$ transition associated with an $A_3^+$ soft mode in which the NiO$_4$ squares undergo antiphase in-plane rotations \cite{cano21b,chen21-instability}.

If the film relaxes, epitaxial strain will typically decrease with the distance to the interface and the lattice parameters should eventually recover their bulk values as the thickness of the film increases. This overall relaxation should occur following a power-law behavior ---rather than an exponential one--- due to the long-range nature of the strain field. This has been quantified for (Nd,Sr)NiO$_2$/SrTiO$_3$ in \cite{ariando21-interfacial}. 
Specifically, the $c$ lattice parameter is found to decrease from 3.42~{\AA} for their 5.1~nm thick sample to 3.36~{\AA} whenever the thickness is larger than 7~nm. 
At the same time, the superconducting $T_c$ is found to correlate with such a decrease in the strain as it displays an increase from 6~K to 13~K.  
On the other hand, the $T_c$ has been reported to increase in (Pr,Sr)NiO$_2$ grown on (LaAlO$_3$)$_{0.3}$(Sr$_2$AlTaO$_6$)$_{0.7}$ (LSAT) as compared to the original SrTiO$_3$ substrate, for 8~nm-thick thin films in both cases \cite{zhu21-substrates}. This increase has been interpreted as due to the additional strain induced by the LSAT substrate. However, in contrast to the above, a higher degree of in-plane compressive (out-of-plane tensile) strain would seem to enhance the $T_c$ in this case. 
%\red{non chiaro} 

The apparent contradiction between these observations may be resolved due to different factors. 
\blue{Two} of them may be just \blue{stoichiometry and sample quality}.
\blue{
The $R$:Ni flux ratio used during the growth of the perovskite films, for example, has been reported to be an important factor for their subsequent reduction into the infinite-layer phase and hence for the observation superconductivity \cite{nie21-fp}.
Beyond that}, samples synthesized in nominally similar conditions can easily show substantial variations in the superconducting $T_c$ and even stay non-superconducting \cite{hwang19a}. 
\blue{
In this respect, the recent report of superconductivity in (La,Sr)NiO$_2$ emphasizes the importance of sufficiently low disorder and high crystallinity \cite{hwang21-La,ariando21-La}.}
Consequently, the quantification of the $T_c$ as a function of the structure may be taken with a grain of salt 
(\blue{in the sense that} the actual error bars may be quite large). 

\section{Chemical reconstructions}

\subsection{Topotatic hydrogen}

Another factor that convolutes with the above is the composition itself. As mentioned above, infinite-layer nickelates are synthesized by topotactic reduction of perovskite precursors using reducing agents such as CaH$_2$. This process, however, may result in the formation of oxide-hydrides $R$NiO$_2$H. 
This possibility has been argued to correlate with both the $R$ element and epitaxial strain \cite{held20topotaticH,zunger21,cano21b}. Specifically, while the parent phases are prone to topotactically incorporate H, both $R$-element substitution and compressive epitaxial strain can limit this possibility. The latter is consistent with the transformation of the infinite-layer NdNiO$_2$ phase into a fluorite-defect structure NdNiO$_x$H$_y$ as a function of the distance to the interface with (001) SrTiO$_3$ reported in \cite{hasegawa16-dalton}.

\begin{figure*}[t!]
    \includegraphics[width=.85\textwidth]{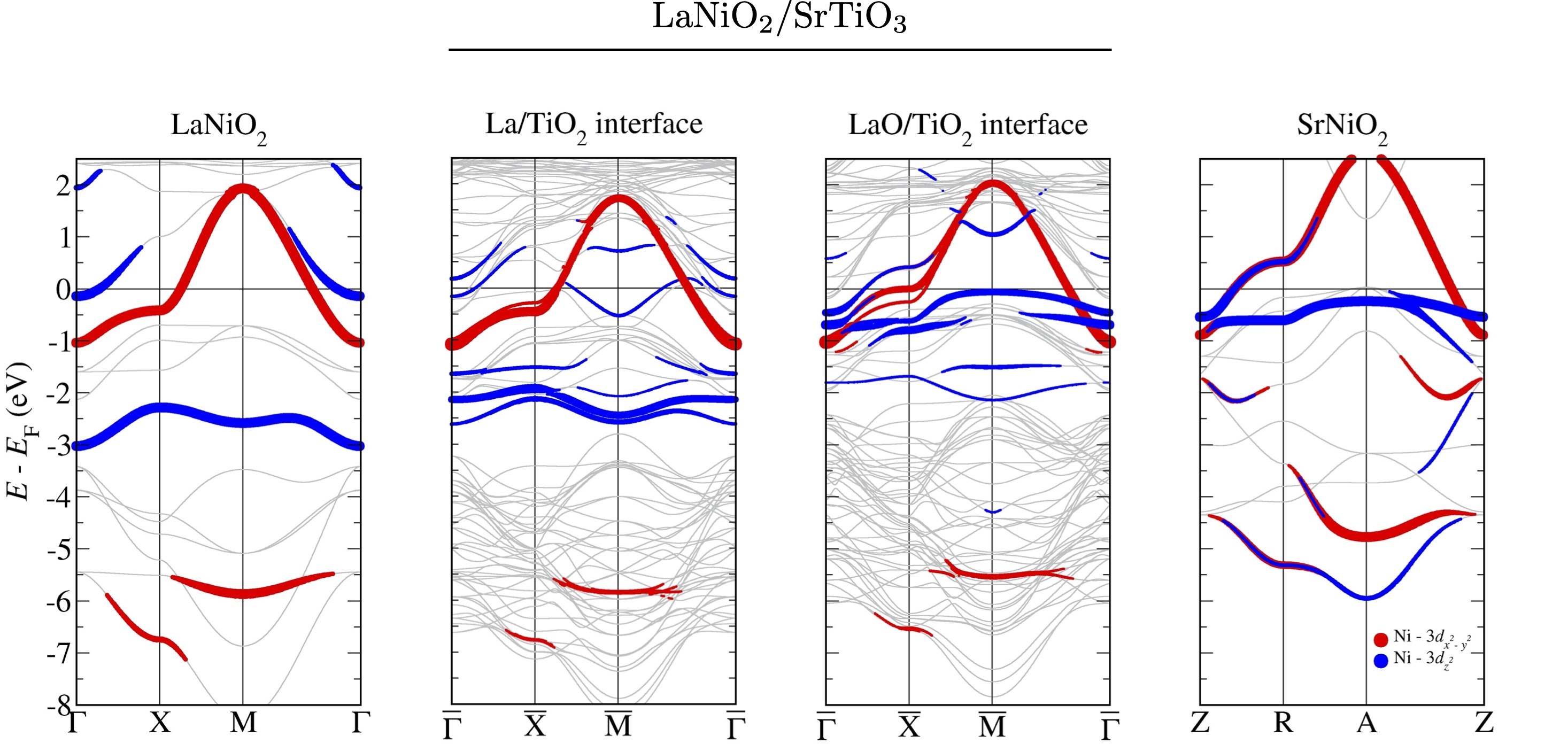}
    \caption{
    Electronic band structure of the LaNiO$_2$/SrTiO$_3$ heterostructure for different interfacial configurations compared to bulk LaNiO$_2$ and SrNiO$_2$ \blue{with the same crystal structure and lattice parameters} (the $\Gamma$-X-M-$\Gamma$ and Z-R-A-Z paths fold to $\bar \Gamma$-$\rm \bar X$-$\rm \bar M$-$\bar \Gamma$). The presence of interfacial apical oxygens introduces an effective hole doping that locally changes the occupation of the Ni-3$d$ states, so that oxidation state of the Ni atom increases from +1 towards +2. Adapted from \cite{cano20c}.
    }
    \label{fig:local-bands}
\end{figure*}

\subsection{Apical oxygens} 

Beyond that, even without insertion of topotactic hydrogen, the reduction process can be different at the interface and further away from it and may result, in particular, in the presence of apical oxygens.
%, and also in different possibilities for the corresponding terminations. 
This question was first addressed by means of DFT calculations in \cite{cano20c,pentcheva20a}. Both these works predicted the presence of these interfacial apical oxygens, which have recently been confirmed in an {\it ad hoc} experimental study supplemented with DMFT calculations \cite{benckiser21-prb}.

The presence of apical oxygens at the interfaces can be seen as a sort of `chemical' reconstruction of the nickelate that locally introduces hole doping. This is illustrated in Fig. \ref{fig:local-bands}. 
Without reconstruction, the characteristic band structure of the bulk would be essentially preserved at the interface. Accordingly, the Ni atom would display the same oxidation state everywhere. 
However, if the apical oxygen remains at the interface, then the local band structure changes towards the $R$NiO$_3$ ---or rather $R$NiO$_{2.5}$--- case. This implies that the Ni oxidation state locally approaches +2 rather than +1. This is a very interesting possibility since it may emulate the hole doping obtained by means of the $R \to$~Sr or Ca substitution, necessary to further promote superconductivity.  
However, the additional charge introduced in this way may be distributed over a distance in the film. 
Thus, in the ultrathin limit, the effective degree of doping would eventually be controlled by the thickness of the film. 
In fact, a gradual increase of the Ni oxidation state as the thickness of the superconducting films decreases has been inferred from XAS data in \cite{ariando21-interfacial}. 
These carriers, however, have been argued to be self-trapped at the interfaces due to Hund's coupling according to DMFT calculations \cite{benckiser21-prb}. 
We note that this may be reminiscent of the metal-insulator transition that takes place in the perovskite counterparts. At the same time, carriers from $B$-site atom of the substrate ---Ti in SrTiO$_3$ vs Ga in LaGaO$_3$ for example--- also participate in the overall process so that the self-trapping may be surface specific. 

\subsection{Additional thin film/substrate terminations}

\begin{figure}[t!]
    \includegraphics[width=.333\textwidth]{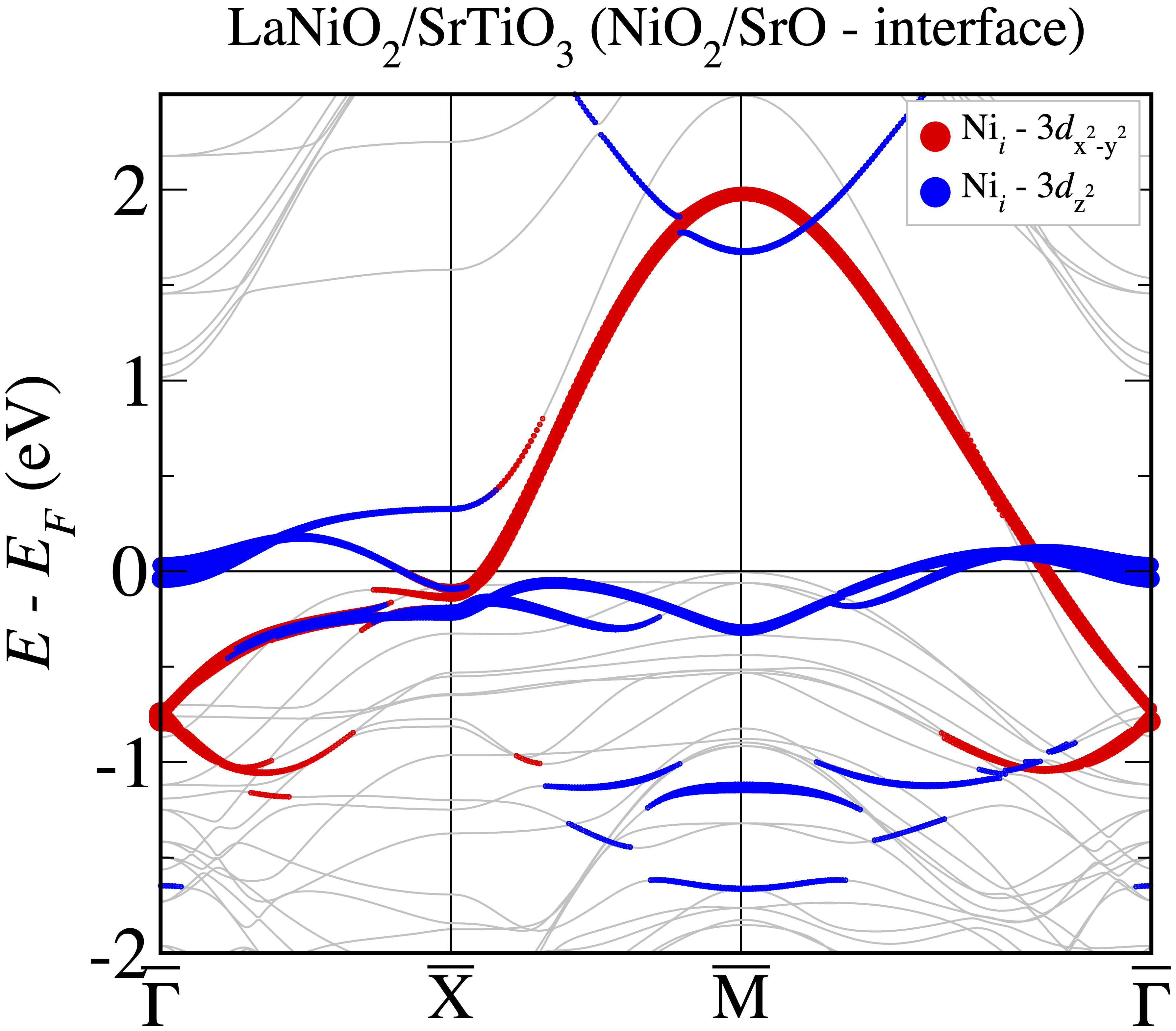}
    \caption{Electronic band structure of the LaNiO$_2$/SrTiO$_3$ heterostructure for the NiO$_2$/SrO interfacial configuration (the colors highlight the main contributions of the interfacial Ni-3$d$ states near the Fermi level and a $3/3$ superlattice is considered). The effective interfacial doping for this configuration depletes the self-doping from the $R$-5$d$ states characteristic of the bulk and places the Ni-3$d_{z^2}$ states at the Fermi level introducing flatband features. Adapted from \cite{cano20c}.    
    }
    \label{fig:local-bands2}
\end{figure}

The situation may be even more complex since, depending on the actual synthesis conditions, additional chemical reconstructions have been shown to be possible at the interface \cite{cano20c}. 
In particular, the most likely $R$O/TiO$_2$ configuration may be replaced by the NiO$_2$/SrO one. That is, the termination may be different. In that case, there would be an extra local doping associated to the interfacial Sr, superimposed to the one from the apical oxygen. 
In fact, the interfacial cell would be La$_{0.5}$Sr$_{0.5}$NiO$_{2.5}$ so that the oxidation state of the Ni atom would increase to $+$2.5.
This overdoping again may be distributed over a distance so that the actual occupation of the Ni-3$d$ states may be determined by the thickness of the film. 
In the ultrathin limit, in particular, the corresponding electronic structure has been shown to undergo drastic changes (see Fig. \ref{fig:local-bands2}). 
Specifically, the initial self-doping effect provided by the $R$-5$d$ states in the bulk will be completely depleted by the local SrO overdoping at the interface. At the same time, the intefacial Ni-3$d_{z^2}$ states are driven even closer to the Fermi energy so that they will manifestly participate in the low-energy physics. The Ni-$e_g$ sector then can be expected to be fully active, and supplemented with a markedly flatband character of the interfacial Ni-3$d_{z^2}$ states. 
In addition, the charge-transfer energy also decreases. As a result of these local changes, both Hubbard's and Hund's couplings may be equally important for quantifying the corresponding correlation effects and/or distinct interface-specific instabilities may appear.  

\subsection{Surface}

The above changes with respect to the bulk may also happen in thin films with asymmetric boundaries where the effective built-in electric field that appears due to the corresponding polar discontinuities play a role. 
Specifically, the screening of this field implies a charge transfer that adds to the aforementioned effects. 
%in determining the resulting structural and electronic reconstructions. 
In the case of asymmetric boundaries, polar layers can be formed at both the surface and the interface \cite{dagotto20,pentcheva21a}. 
These layers display antiparallel NiO$_2$ displacements, but otherwise are decoupled. 
Further, the aforementioned depletion of the $R$-5$d$ states can also be obtained at the surface while a two-dimensional electron gas extending over several layers can be formed at the interface  \cite{dagotto20,pentcheva20a}. 
Besides, the combined effect of magnetism ---$G$-type antiferromagnetic order--- and correlations at the DFT+$U$ level has been found to enhance the itineracy of the 
Ni-3$d_{z^2}$ orbitals at the interface with the substrate, while the magnetism is essentially suppressed at the surface to vacuum \cite{dagotto20}. 

The focus has been put on the surface properties in \cite{thomale20b} taking into account the possibility of apical oxygens anticipated in \cite{cano20c,pentcheva20a}. 
Thus, it has been confirmed that different terminations yield different electronic structures also at the surface (see Fig. \ref{fig:local-bands}). 
This is further shown to modify qualitatively the corresponding Fermi surface. As a result of this modification, it is argued that the $d$-wave superconducting gap expected for the bulk may transform into a $s_\pm$-wave one at the NiO$_2$-terminated surface. This provides a rather natural explanation to the local changes in the tunneling spectrum observed in \cite{wen20a}. Further, it suggests that a surface $s+id$-wave state may also be realized under the appropriate conditions.

\section{Outlook}

The thin-film approach has proven its advantages to overcome the thermodynamic fragility of the infinite-layer nickelates, thereby demonstrating the emergence of unconventional superconductivity in these systems. 
Besides, this approach offers additional degrees of flexibility to further investigating this phenomenon. 

Epitaxial strain can be used not only to fine tune the electronic structure of these nickelates and hence their ``cuprateness'', but also the underlying crystal structure itself.  
Beyond that, the presence of apical oxygens at the interfaces/surfaces enables a local control of the effective doping without the need of rare-earth-element substitution. This apical-oxygen doping can be further supplemented with an extra contribution from the interfacial terminations themselves. 
This may be useful to bypass the sample-quality issues associated with doping at the rare-earth site. In fact, the overall local doping obtained in that way may be a particularly effective control parameter in the ultrathin limit.  

Another route yet to be explored experimentally is the engineering of the electronic properties via nickelate-based superlattices. This idea was put forward some time ago with the perovskite phases as building blocks \cite{held09-prl}. 
This concept can now be supplemented with the possibility reducing these phases, thereby mimicking higher-order layered nickelates. 
The ($R$NiO$_2$)$_3$/(SrTiO$_3$)$_3$ superlattices, for example, have been shown to display many analogies with the trilayer systems $R_4$Ni$_3$O$_8$ [or $R$O$_2$($R$NiO$_2$)$_3$] \cite{cano20c}. Conceptually, the superconducting pentalayer nickelate can also be simulated in this way. This meta-material approach may enable to bypass the thermodynamic instability of these higher-order phases and, at the same time, exploit the interfaces ---rather than the spacers of layered bulk materials--- as reservoirs for charge doping. 
\blue{In addition, it can also help to clarify the interplay between superconductivity and other degrees of freedom. Charge density waves, in particular, have very recently been reported to display an intriguing dependence on the presence of capping layers in infinite-layer nickelate thin films calling for further investigations \cite{lee21-cdw,preziosi21-cdw,zhou21-cdw}. }

\acknowledgements{%L.I. acknowledges support from the Région Île-de-France through the DOPNICKS project. 
L.I. acknowledges the support and funding from the Île de France region and the European Union's Horizon 2020 research and innovation programme under the Marie Sklodowska-Curie grant agreement Nº21004513 (DOPNICKS project).
A.C. was supported by ANR, Grant ANR-18-CE30-0018.}

\bibliography{bib.bib}

\end{document}